\documentclass[prb,aps,amssymb,nofootinbib,twocolumn,showpacs]{revtex4}
\usepackage{amsmath}
\usepackage{amssymb}
\usepackage{amsthm}
\usepackage{amsfonts}
\usepackage{algorithmic}
\usepackage{enumerate}
\usepackage{latexsym}
\usepackage[dvips]{graphicx}

\newcommand{\beq}{\begin{equation}}
\newcommand{\eneq}{\end{equation}}
\newcommand{\beqs}{\begin{equation*}}
\newcommand{\eneqs}{\end{equation*}}

\input{epsf}

\begin{document}

\tolerance 10000

\title{Quantum States of Matter of Simple Bosonic Systems: \\ BEC's, 
Superfluids and Quantum Solids}

\author { Zaira Nazario$^\dagger$ and
David I. Santiago$^{\dagger, \star}$ }

\affiliation{ $\dagger$ Department of Physics, Stanford
University,Stanford, California 94305 \\ 
$\star$ Gravity Probe B Relativity Mission, Stanford, California 94305}

\begin{abstract}
\begin{center}

\parbox{14cm}{ The phase diagram of a single component Bose system in a 
lattice at zero temperature is obtained. We calculate the variational energies 
for the Mott insulating and superfluid phases. Below a certain critical 
density the Mott insulating phase is stable over the superfluid phase for low 
enough tunelling amplitude regardless of whether the number of bosons is or is 
not incommensurate with the lattice. The transition is discontinuous as the 
superfluid order parameter jumps from a finite value to zero at the 
Mott transition.}

\end{center}
\end{abstract}


\date{\today}

\maketitle

The purpose of the present work is to study the phase diagram of a single 
component Bose gas at zero temperature. Specifically, we have in mind
a system of bosons in an optical lattice, where the wells on each site are 
made deeper thus causing a transition from the superfluid phase into a phase
where bosons are localized on the lattice sites, i.e a quantum solid or Mott 
insulating phase\cite{ib,fisher}. Recently such a transition has been 
observed experimentally\cite{ib}.

The original theoretical work by M.P.A. Fisher {\it et. al.}\cite{fisher} on 
the superfluid-Mott transition in
bosonic systems predated the experimental realization of artificially 
engineered Bose-Einstein condensates\cite{beca,becb,becc}(BEC's), which are 
superfluid systems as they posses a finite sound speed\cite{bec2,landau,bog}
and suppressed long-wavelength scattering\cite{bec3,bec4}. 
In that original work the transition was asserted to be continuous and possible
only for a number of bosons commensurate with the lattice, i.e. integral number
of bosons per site. While widely believed to be true, both assertions are
quite puzzling and may contradict experiment\cite{ari}. In fact, recent 
theoretical\cite{glass} and numerical work\cite{mc} has pointed to a 
transition that is incommensurate and ``nonstandard'' even for a pure system
\cite{mc}. In the present note we extend the work in the pure system within
the Bogolyubov approximation for the superfluid\cite{bog,bog2}

The superfluid-Mott transition is between a superfluid and a quantum solid. In 
systems without disorder, fluid-solid transitions are usually discontinuous as 
the order parameters are too different\cite{landau2}. In the superfluid system 
that is certainly the case.  The Mott insulating phase is characterized by a 
well defined number of bosons per site, one could take the density as its 
order parameter. The superfluid phase has off-diagonal order with breaking 
of gauge invariance\cite{phil}, i.e. $U(1)$, characterized by a coherent 
ground state of Bogolyubov pairs\cite{bog}. The transition will be 
characterized by a {\it discontinuous} jump in the superfluid 
response\cite{ari}.

We calculate the variational energies of Mott insulating and superfluid phases 
with both
{\it commensurate} and {\it incommensurate} number of bosons. We find that
below some critical value $r_c$ of the ratio $t/U$ the Mott insulating phase is
energetically favorable over the superfluid phase as long as the number of 
bosons per site is not too high, i.e. less than a critical value $n_c$. Below 
$r_c$ and $n_c$ the superfluid phase is {\it never} stable {\it regardless} of 
commensurability.

\begin{figure}[ht]
\centering
\resizebox{6cm}{!}{%
   \includegraphics*{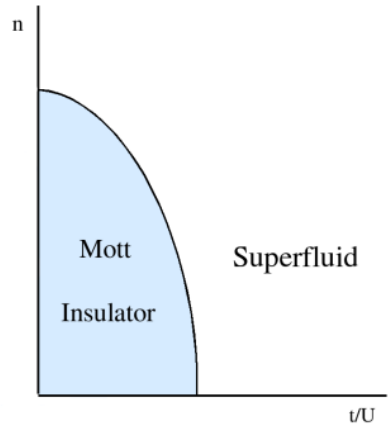}}
\caption{Schematic Phase Diagram of a single component Bose system at zero 
temperature. n is the average number of bosons per site.}
\end{figure}

When the number of bosons per site is higher than the critical number $n_c$,
the quantum solid phase is never stable {\it within our approximations} and 
there is only a superfluid phase even for arbitrarily small $t/U$. On the other
hand, we believe our approximations, which are apt at low number of bosons per 
site, {\it fail} at high densities and there will be 
``crossover'' to physics that is more like that of Josephson coupled 
arrays\cite{fisher,ja}. This is a matter for further study. The phase diagram 
is shown in figure 1.

A very apt prototype Hamiltonian to study bosons in a lattice (which, of 
course can be an optical lattice) is the Bose Hubbard Hamiltonian:

\begin{multline} \label{ham}
\mathcal H = \frac{-t}{2} \sum_{ <ij> } ( a_i^\dagger a_j + a_j^\dagger a_i ) 
+ U \sum_i a_i^\dagger a_i ( a_i^\dagger a_i - 1 ) \\
 - \varepsilon \sum_i 
a_i^\dagger a_i
\end{multline}

\noindent where $t$ is the nearest-neighbor tunneling or hopping amplitude, $U$
is the on-site repulsion, and $\varepsilon$ is the well depth for each lattice 
site. We will suppose our lattice to have $N$ sites and our system to have 
$M$ bosons with $M/N$ no necessarily an integer. 

When $U=0$, the ground state of such a Hamiltonian is a BEC with all atoms in 
the zero momentum state. Such a system is not superfluid as its excitation 
spectrum has quadratic dispersion\cite{landau}. An arbitrary small amount of
repulsion causes the system to order and become superfluid. The ground
state, first found by Bogolyubov\cite{bog}, is 

\beq \label{bogow}
|\psi_0 \rangle = \prod_{\bf k} 
\frac{ 1 }{ u_{\bf k} } e^{ -(v_{\bf k} / u_{\bf k}) a_{\bf k}^\dagger 
a_{-\bf k}^\dagger} \; |0 \rangle
\eneq

\noindent where 

\beq \label{uv}
u_{\bf k} = \frac{ 1 }{ \sqrt{2} } \sqrt{ \frac{ \tilde{\epsilon_{\bf k}} }
{ E_{\bf k} } + 1 } \qquad v_{\bf k} = \frac{ 1 }{ \sqrt{2} } 
\sqrt{ \frac{ \tilde{\epsilon_{\bf k}} }{ E_{\bf k} } - 1 }
\eneq

\noindent with $\tilde { \epsilon_{\bf k} } = \epsilon_{\bf k} + 
\frac{2 M_0}{N} U$, where $M_0$ is the number of particles in the condensate, 
and $\epsilon_{\bf k} = - 2 t ( \cos{ k_x a } + \cos{ k_y a } + \cos{ k_z a } )
+ 6 t$ is the dispersion in the lattice which goes like $t a^2 k^2 = \hbar^2 
k^2 / 2 m$ in the long wavelength limit. In order to perform our
calculations we will introduce a cut-off defined by the point at which
the kinetic energy becomes comparable to the on-site repulsion, i.e. 
$k_c = \frac{\sqrt{3} \pi}{a} \sqrt{ \frac{U M_0}{t N} }$. 

In the Bogolyubov state, ``conservation'' of bosons imposes the sum rule

\begin{align} \label{sum}
\nonumber M &= \sum_{\mathbf k \neq 0} \langle a_{\bf k}^\dagger a_{\bf k} 
\rangle  + M_0 = \sum_{\mathbf k \neq 0} v_{\bf k}^2 + M_0 \\
\nonumber 1 &= \frac{3}{16} a^3 k_c^3 \frac{N}{M} + M_0 / M \\
1 &= \frac{3}{16} \pi^3 \left( 
\frac{3 M_0 U}{N t} \right)^{3/2} \frac{N}{M} + \frac{M_0}{M} 
\end{align}

\noindent which determines the number of particles in the condensate. In
the Bogolyubov state we have a superfluid order parameter which corresponds to 
boson pair correlations

\begin{align} 
\nonumber \langle a_{\bf k} a_{ - {\bf k} } \rangle 
&= \frac{ - v_{\bf k} }{ u_{\bf k} } \\
\nonumber &= - \sqrt{ \frac{ (\tilde \epsilon_{\bf k} / E_{\bf k}) - 1 }
{\tilde \epsilon_{\bf k} / E_{\bf k}) + 1} } \\
&= - \sqrt{ \frac{ \epsilon_{\bf k} 
+ \frac{2 M_0 U}{N} - E_{\bf k} }{ \epsilon_{\bf k} + \frac{2 M_0 U}{N} 
+ E_{\bf k} } }
\end{align}

\noindent with $E_{\bf k} = \sqrt{\epsilon_{\bf k}}\sqrt{ \epsilon_{\bf k} 
+ 4 \frac{M_0}{N} U}$ the quasiparticle excitation spectrum of the superfluid. 
Since at
long-wavelengths it is a sound spectrum, the system superflows by the
Landau argument\cite{landau}. Even though a Bose condensate is necessary for 
the Bose system to be a quantum liquid, the Bogolyubov pairs arising from the 
repulsion between atoms give the system the necessary rigidity to be 
superfluid. Therefore the superfluid order parameter is the amplitude of 
Bogolyubov pairs as whenever this amplitude vanishes, the system dissipates. 

\begin{figure}[ht]
\centering
\resizebox{8cm}{!}{%
   \includegraphics*{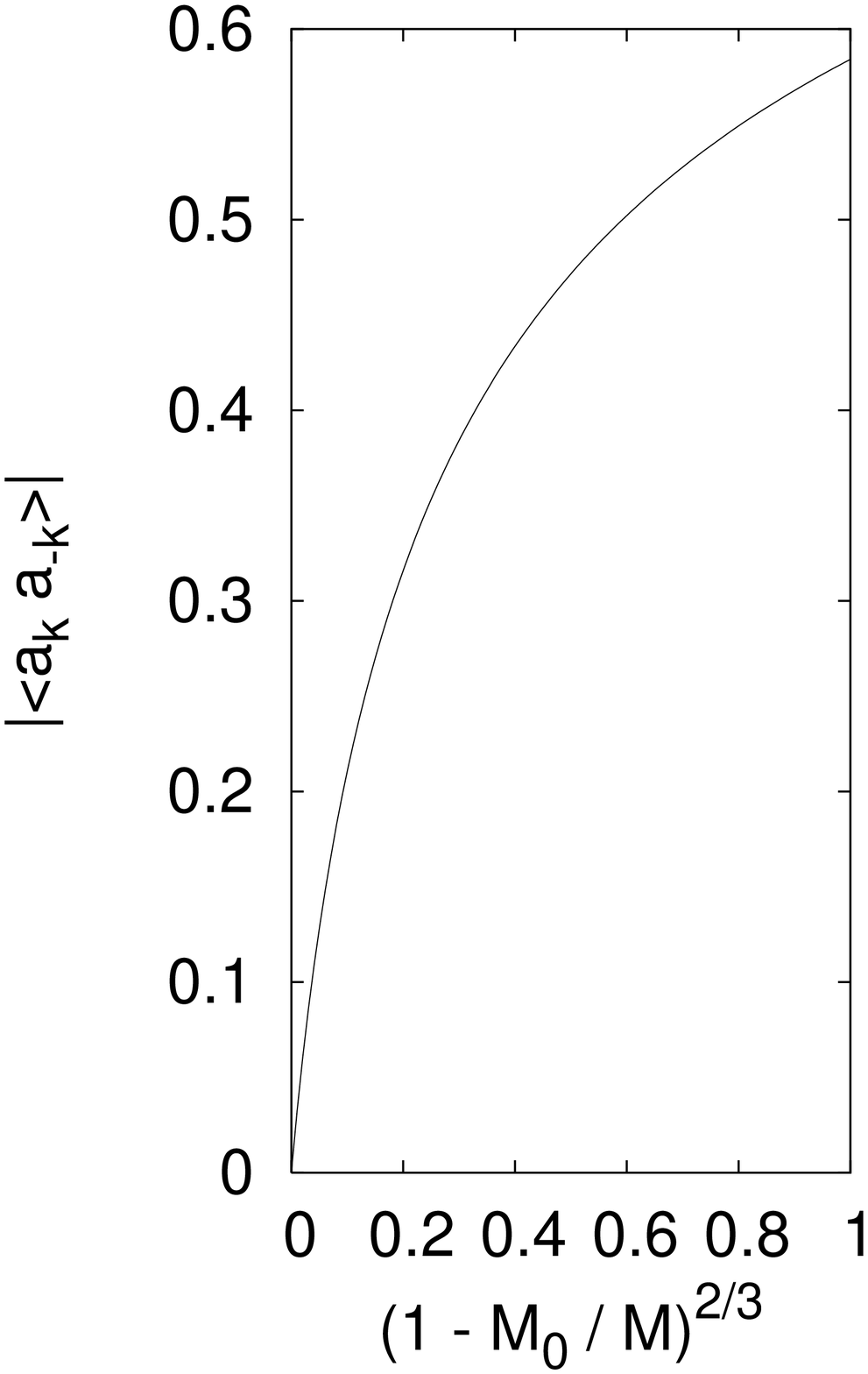}\includegraphics*{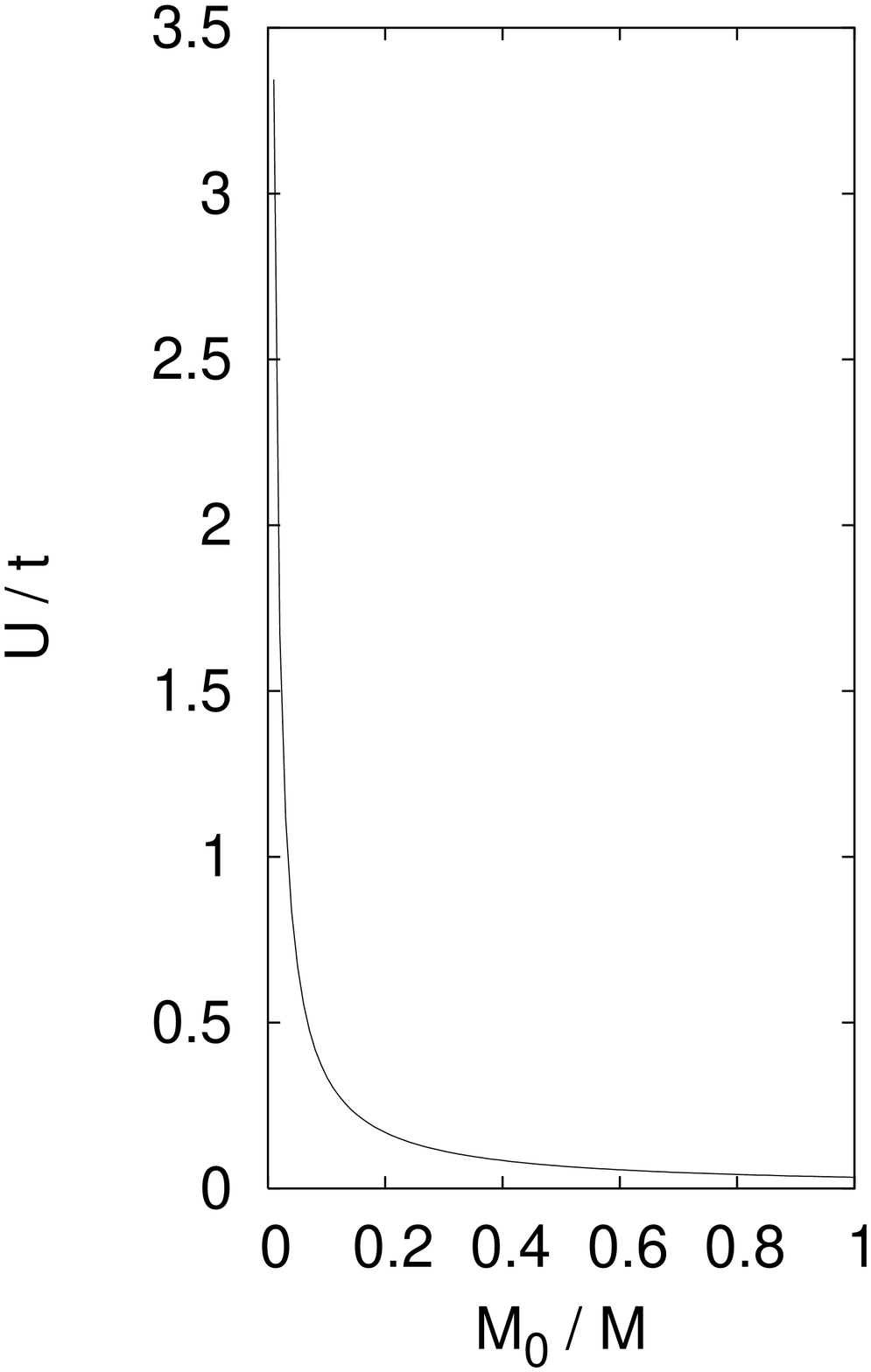}}
\caption{(a) Dependence of the order parameter on the fraction of bosons
in the condensate, (b) $U / t$ vs. $M_0 / M$}
\end{figure}

In BEC's in an optical lattice the Mott transition is induced by making the 
wells on the lattice sites deeper ($\varepsilon$ is the lattice depth) thus 
suppressing the tunelling amplitude $t$ 
between near neighbor sites, i.e. making the bosons ``heavy''. Thus the 
relevant ratio of on-site repulsion $U$ to hopping $t$ increases. As this
ratio increases, the superfluid order parameter increases as shown in the
figure 2. The increase of the order parameter means that the superfluid
phase is becoming more stable over the Bose condensed gas. On the other hand, 
we also plot the number of particles in the condensate. This number is becoming
small as $U/t$ increases. As the condensate gets depleted the system is 
becoming ``less fluid'' and will solidify into a Mott insulating phase with a 
well defined number of bosons per site. Since the Bogolyubov order parameter 
will be zero in the quantum solid phase, there will be a {\it discontinuous} 
jump in the superfluid response at the Mott transition.

The Mott-insulating ground state wavefunction for $p N \le M < (p + 1) N$, 
with $p$, an integer is

\beq \label{M+}
|\psi_1 \rangle = \prod_{j = 1}^{M-pN} a_{i_j}^\dagger 
\prod_{l = 1}^N (a_l^\dagger)^p | 0 \rangle
\eneq

\noindent where $a_{i_j}$ and $a_l^\dagger$ are boson creation operators at
lattice sites $i_j$ and $l$ respectively. For an incommensurate system, 
$p N < M < (p + 1) N$, the ground state will not be strictly periodic as there 
are $M - pN$ that do not fill all lattice
sites. No two of these extra bosons would occupy the same site when the system 
is in its ground state, but otherwise they will be randomly distributed. In 
real life, besides on-site interactions, there are interactions
acting at long distances. They can be attractive or repulsive. These
interactions are weaker than the on site interaction $U$, hence not
important for a commensurate system. But if there is incommensuration,
these interactions will probably play a role in determining the
positions of the extra bosons.

Let us take a look at the expectation value of the superfluid order parameter 
in the Mott phase:

\beq
\langle a_{\bf k} a_{ - {\bf k} } \rangle = \frac{ 1 }{ N } \sum_{<ij>} 
e^{ i {\bf k} \cdot ( {\bf r}_i - {\bf r}_j ) } \langle \psi_1 | a_i a_j | 
\psi_1 \rangle
\eneq

\noindent where

\beq
\langle 0 | \prod_l a_l^{q_l} a_i a_j 
\prod_l (a_l^\dagger)^{q_l} | 0 \rangle 
= \prod_{l \neq i, j} \langle 0 | F_{ij} a_l^{q_l} 
(a_l^\dagger)^{q_l} | 0 \rangle 
\eneq

\noindent In the expression above, $q_l$ indicates the number of
bosons in site $l$ and $F_{ij}$ is a function of bosonic operators on
sites $i$ and $j$. 

\begin{align}
\nonumber \prod_{l \neq i, j} \langle 0 | F_{ij} a_l^{q_l} 
(a_l^\dagger)^{q_l} | 0 \rangle 
&= \prod_{l \neq i, j} \langle 0 | F_{ij} a_l^{q_l - 1} 
a_l (a_l^\dagger)^{q_l} | 0 \rangle \\
\nonumber &= \prod_{l \neq i, j} \langle 0 | F_{ij} a_l^{q_l - 1} 
q_l (a_l^\dagger)^{q_l - 1} | 0 \rangle \\
\nonumber & + \prod_{l \neq i, j} \langle 0 | F_{ij} a_l^{q_l - 1} 
(a_l^\dagger)^{q_l} a_l | 0 \rangle \\
&= \prod_{l \neq i, j} \langle 0 | F_{ij} a_l^{q_l - 1} 
q_l (a_l^\dagger)^{q_l - 1} | 0 \rangle 
\end{align}

\noindent Continuing the process we find

\begin{align}
\nonumber \prod_{l \neq i, j} \langle 0 | F_{ij} & a_l^{q_l - 1} 
q_l (a_l^\dagger)^{q_l - 1} | 0 \rangle 
= \prod_{l \neq i, j} q_l ! \langle 0 | F_{ij} | 0 \rangle \\
&= \prod_{l \neq i, j} q_l ! \langle 0 | a_i^{q_i} a_j^{q_j} 
a_i a_j (a_i^\dagger)^{q_i} (a_j^\dagger)^{q_j} | 0 \rangle 
\end{align}

\noindent Since there is a non matching number of operators acting on
the ground state this yields 0. The Bogolyubov order parameter is 0 in the Mott
phase

\beq
\langle a_{\bf k} a_{ - {\bf k} } \rangle = 0
\eneq

\noindent Actually this is obvious because the Mott state has 
a well defined number of bosons per site and number phase uncertainty makes the
off-diagonal order zero. We thus see that, as anticipated, the only way to go 
into the Mott insulating phase is to have a discontinuous change in the 
superfluid order parameter. The Mott transition is discontinuous, i.e. first 
order, with the superfluid order parameter, and thus the superfluid response, 
experiencing a sudden jump to 0. 

The transition into the Mott insulating phase will occur when the
Mott wavefunction (\ref{M+}) has less variational energy than the Bogolyubov
superfluid wavefunction (\ref{bogow})
With the Hamiltonian given by (\ref{ham}), the energy of
the Mott state is found to be

\begin{multline}
\langle \psi_1 | \mathcal H | \psi_1 \rangle = -t \sum_{ i, \delta } 
\langle \psi_1 | a_{i + \delta}^\dagger a_i | \psi_1 \rangle \\
- (\varepsilon + U) \sum_i \langle \psi_1 | n_i | \psi_1 \rangle 
+ U \sum_i \langle \psi_1 | n_i^2 | \psi_1 \rangle 
\end{multline}

\noindent The first term corresponds to the kinetic energy. It
destroys a boson in site $i$ and creates it in its nearest neighbor
site $i + \delta$. Thus the lack of overlap of the resulting wavefunction
with the ground state gives 0. We get

\begin{multline}
\nonumber \langle \psi_1 | \mathcal H | \psi_1 \rangle 
= - (\varepsilon + U) M 
+ U (M - p N) (p + 1)^2 \\
+ U [N - (M - p N)] p^2 \\
\end{multline}

\beq \label{E+}
\langle \psi_1 | \mathcal H | \psi_1 \rangle 
= - \varepsilon M + 2 p U M - (p + 1) p N U 
\eneq

\noindent We see that $2 p U M \ge (p + 1) p N U$. After plenty of tedious 
algebra, with our cut-off method the variational energy the superfluid ground 
state can be estimated to be

\begin{align}
\nonumber \langle \psi _0| \mathcal H |\psi_0 \rangle \simeq - \varepsilon  M 
&+ \frac{3}{32} \pi^5 \left ( \frac{3 M_0 U}{t N} \right )^{5/2} N t \\
\nonumber &+ U  N \left ( \frac{3 M_0 U}{N t} 
\right )^{3} \left ( \frac{ 1 }{ 27 } 
+ \frac{ \pi^3 }{ 8 \sqrt{3} } 
+ \frac{ 27 \pi^6 }{ 16^2 } \right ) \\
&- 6 t M
\end{align}

\noindent For $U/t$ small, we see that the interaction terms in the superfluid 
are almost irrelevant when compared with those of the Mott phase so the
superfluid state is the stable one. We also notice that for not too large an 
$M$ the interaction terms in the superfluid grow a lot faster than the 
interaction terms in the Mott insulating phase. Hence for some critical $U/t$
the Mott insulating phase will become stable causing the system to undergo a 
quantum phase transition. Whether $M$ is commensurate or incommensurate with 
the lattice is irrelevant to the energetics: The state with {\it all} bosons
localized is energetically favorable to the state with {\it all} bosons 
superfluid. Whether the instability of the superfluid is toward a phase with
all bosons localized, or into a state with coexistence of a commensurate Mott 
insulator with the leftover bosons superfluid is not straightforward to answer
variationally. It can be answered experimentally as in the case of coexistence
the superfluid response will have a discontinuous jump to a smaller 
{\it nonzero} value, while in the case of all bosons localized the superfluid 
response will have a discontinuous jump to 0.

The phase diagram of the simple Bose system at zero temperature shown in 
figure 1 differs considerably from the one proposed in 
the pioneering theoretical work on Bose systems\cite{fisher} and thus requires
some comments. The first and, perhaps, most important difference is that
we do not consider a phase only model as it is not appropriate to the Mott 
insulator at low densities. In that original work, the phase of the superfluid 
was disordered by
the increasing repulsion thus leading to a transition. In such a transition the
solid would have a supefluid order parameter within {\it each} well, but the 
phase of the order parameter will have become randomized exactly analogous to 
what happens in Josephson junction arrays when the charging energy is 
sufficiently large. Such an insulating phase does not correspond to the Mott 
insulator we studied here and cannot exist at small number of bosons per site 
as one needs a macroscopic number of particles to have a superfluid order 
parameter. Finally, we stress the the phase only model studied in the early 
work\cite{fisher,ja} is a correct model of an array of Josephson coupled
superfluid systems and should work for bosons in an optical lattice when the 
number of bosons per site is large enough to make superfluid order within 
each well possible. Josephson coupled systems can easily 
be studied in an optical lattice\cite{mark1,mark2}. It will be extremely 
interesting to see what the experimental phase diagram turns out to be.

The correct physics for the Bose system phase diagram can be easily 
differentiated experimentally as the original work would predict that the 
superfluid response goes continuously to zero at the transition. We predict 
that the superfluid response will have a sharp jump to zero at the transition. 
We predict a Mott transition irrespective of commensurability as long as the 
number of bosons per site is less than some critical density $n_c$. At high 
densities there should be a ``crossover'' into the physics of coupled 
Josephson arrays\cite{fisher,ja}.

\noindent {\bf Acknowledgements:} We thank Ari Tuchman and Mark Kasevich
for numerous and stimulating discussions, and for the extreme generosity with 
their data prior to publication. Bob Laughlin and Sandy Fetter provided very 
useful criticism and suggestions. Zaira Nazario was supported by
The School of Humanities and Sciences at Stanford University. 
David I. Santiago was supported by NASA grant NAS 8-39225 to Gravity Probe B.


\begin{thebibliography}{99}
\bibitem{ib} M. Greiner {\it et. al.}, Nature {\bf 415}, 39 (2002).
\bibitem{fisher} M.P.A. Fisher {\it et. al.}, Phys. Rev. B. {\bf 40},
546 (1989).
\bibitem{beca} K. B. Davis {\it et. al.}, Phys. Rev. Lett. {\bf 75}, 3969 
(1995).
\bibitem{becb} C. C. Bradley {\it et. al.}, Phys. Rev. Lett. {\bf 75}, 1687
(1995).
\bibitem{becc} M. H. Anderson {\it et. al.}, Science {\bf 269}, 198 (1995). 
\bibitem{bec2} M. R. Andrews {\it et. al.}, Phys. Rev. Lett. {\bf 79}, 553 
(1997).
\bibitem{bec3} W. Ketterle {\it et. al.}, Phys. Rev. Lett. {\bf 83}, 2876 
(1999).
\bibitem{bec4} A. G\"{o}rlitz {\it et. al.}, Phys. Rev. A {\bf 63}, 041601 
(2001).
\bibitem{glass} B. Damski {\it et. al.}, Phys. Rev. Lett. {\bf 91}, 080403
                (2003).
\bibitem{mc} G. G. Batrouni {\it et. al.}, Phys. Rev. Lett. {\bf 89}, 117203
             (2002).
\bibitem{bog} N. N. Bogolyubov, J. Phys. USSR {\bf 11}, 23 (1947).
\bibitem{bog2} A. M. Ray {\it et. al.}, J. Phys. B {\bf 36}, 825 (2003).
\bibitem{bog3} Van Oosten {\it et. al.}, Phys. Rev. A {\bf 63} 053601 (2001).
\bibitem{landau} L. D. Landau, J. Phys. USSR {\bf 5}, 71 (1941).
\bibitem{ari} Ari Tuchman and Mark Kasevich, private communication of not yet 
published data that has been presented at several conferences. Preliminary 
data indicates the possibility that the transition is indeed first order under 
specific conditions.
\bibitem{ja} Batrouni {\it et. al.}, Phys. Rev. Lett. {\bf 65}, 1765 (1990).
\bibitem{landau2} L. D. Landau and E. M. Lifshitz, {\it Statistical Physics},
\copyright Reed Educational Publishing Ltd 1980. 
\bibitem{phil} P. W. Anderson, Rev. Mod. Phys. {\bf 38}, 298 (1966).   
\bibitem{mark1} B. P. Anderson, M. A. Kasevich, {\it Science} {\bf 282}, 1686 
(1999).
\bibitem{mark2} F. S. Cataliotti et al., {\it Science} {\bf 293}, 843 (2001).
\end{thebibliography}
\end{document}